\documentstyle[prl,aps,preprint]{revtex}

\let\ssection=\section
\renewcommand{\section}{\setcounter{equation}{0}\ssection}

\begin{document}
\title{Fresnel analysis of the wave propagation in nonlinear electrodynamics}
 
\author{Yuri N.\ Obukhov\footnote{On leave from: Department
of Theoretical Physics, Moscow State University, 117234 Moscow,
Russia}}
\address{Instituto de F\'{\i}sica Te\'orica, UNESP, Rua Pamplona 145,
01405-900 S\~ao Paulo, SP, Brazil}
\author{Guillermo F. Rubilar}
\address{Institute for Theoretical Physics, University of Cologne, 
50923 K\"oln, Germany}

\maketitle
 
\begin{abstract}
We study the wave propagation in nonlinear electrodynamical models. 
Particular attention is paid to the derivation and the analysis of the 
Fresnel equation for the wave covectors. For the class of general nonlinear
Lagrangian models, we demonstrate how the originally quartic Fresnel 
equation factorizes, yielding the generic birefringence effect. We show
that the closure of the effective constitutive (or jump) tensor is necessary 
and sufficient for the absence of birefringence, i.e., for the existence
of a unique light cone structure. As another application of the Fresnel
approach, we analyze the light propagation in a moving isotropic nonlinear 
medium. The corresponding effective constitutive tensor contains non-trivial 
skewon and axion pieces. For nonmagnetic matter, we find that 
birefringence is induced by the nonlinearity, and derive the corresponding 
optical metrics.
\end{abstract}


\section{Introduction}

Wave phenomena belong to the most interesting and important
processes in physics. Among other field theories, nonlinear 
electrodynamics attracts much attention in connection with the 
prominent role played by light in the experimental and theoretical
studies of the structure of spacetime and matter. 

Nonlinearities in electrodynamical models can arise in different ways 
in classical and quantum field theories. For example, the old Born-Infeld 
theory \cite{BI} was a fundamental theory alternative to the classical 
Maxwell electrodynamics which provided a model of a classical electron. 
On the other hand, quantum Maxwell electrodynamics predicts nonlinear 
effects which arise due to the radiative corrections, see 
\cite{HE,Itzykson,heyl}. Finally, in the modern string theories a 
generalized Born-Infeld action naturally arises as the leading part
of the effective string action, see\cite{ark,GiRa,kern}, for example.

Wave propagation in the various nonlinear electrodynamical theories
was studied previously in \cite{blok,lutz,boi,pleb,pleb1} and also
in \cite{nov1,nov2,GiHe,lemos}. A general feature revealed in
these studies is the existence of birefringence. In crystal
optics, the notion of birefringence means the emergence of two
rays (ordinary and extraordinary) with different velocities inside the
material medium. We will use the expression ``birefringence'' in a  
similar sense, associating it with the situation when two 
different light cones exist for the wave normal covectors. However, 
the earlier results are incomplete in the sense that the full Fresnel 
equation, governing the wave normals, was never derived explicitly.
Moreover, it was not demonstrated how it happens that the original 
quartic surface of wave normals reduces to the light cone. That is
the primary interest in our study, and we will try to clarify this 
aspect for the nonlinear electrodynamical models, using and
expanding our earlier results obtained within the framework of linear 
electrodynamics \cite{OFR00,nonsym}. 

Our basic tool will be the {\it general formula} for the Fresnel 
equation derived earlier within linear electrodynamics. Now we observe 
that the analysis of the wave propagation in a general nonlinear model 
reduces to the linear case because the jumps of the derivatives of
the excitation and of the field strength are in all cases related by a 
linear law. We can make then use of our master formula for the Fresnel 
tensor and {\it derive the Fresnel equation for any nonlinear model}, 
and thereby explain the reduction to the light cones.

\section{Electromagnetic waves and Fresnel tensor}

Quite generally, Maxwell's equations for the excitation 2-form 
$H=({\cal D},{\cal H})$ and the field strength 2-form $F=(E,B)$ read
\begin{equation}\label{me}
dH=J, \qquad dF=0 \,.
\end{equation}
Here $J$ is the electric current 3-form. These equations must be 
supplemented by a constitutive law $H = H(F)$.  The latter relation 
contains the crucial information about the underlying physical continuum 
(i.e., spacetime and/or material medium). Mathematically, this constitutive 
law arises either from a suitable phenomenological theory of a medium or 
from the electromagnetic field Lagrangian. It can be a nonlinear or even
nonlocal relation between the electromagnetic excitation and the field
strength.

If local coordinates $x^i$ are given, with $i,j,... =0,1,2,3$, we can
decompose the excitation and field strength 2-forms into their
components according to
\begin{equation}
H = {\frac 1 2}\,H_{ij}\,dx^i\wedge dx^j,\qquad
F = {\frac 1 2}\,F_{ij}\,dx^i\wedge dx^j.\label{geo1}
\end{equation}

We will study the propagation of a discontinuity of the
electromagnetic field following the lines of Ref. \cite{OFR00}, see
also Refs. \cite{nonsym}.  The surface of
discontinuity $S$ is defined locally by a function $\Phi$ such that
$\Phi= const$ on $S$. Across $S$, the geometric Hadamard conditions
are satisfied:
\begin{eqnarray}
&& [F_{ij}] = 0,\qquad [\partial_i F_{jk}] = q_i\, f_{jk}, 
\label{had1}\\
&& [H_{ij}] = 0,\qquad [\partial_i H_{jk}] = q_i\, h_{jk}. 
\label{had2}
\end{eqnarray}
Here $\left[{\cal F}\right](x)$ denotes the discontinuity of a
function ${\cal F}$ across $S$, and $q_i:=\partial_i\Phi$ is the wave
covector. Given the constitutive law $H(F)$, which determines the
excitation in terms of the field strength, the corresponding tensors
$f_{ij}$ and $h_{ij}$, describing the jumps of the derivatives of
field strength and excitation, are related by \cite{dublin}
\begin{equation}\label{kappa}
h_{ij} = {\frac 1 2}\,\kappa_{ij}{}^{kl}\,f_{kl},\qquad {\rm with}
\qquad \kappa_{ij}{}^{kl} := {\frac {\partial H_{ij}}{\partial F_{kl}}}.
\end{equation}
We will call $\kappa_{ij}{}^{kl}$ the {\it jump tensor}. In linear 
electrodynamics, its components coincide with the components of the 
constitutive tensor (which describes the linear law $H_{ij} = {\frac 1 2}
\,\kappa_{ij}{}^{kl}\,F_{kl}$), and they are independent of the 
electromagnetic field. However, in general the jump tensor
$\kappa_{ij}{}^{kl}$ is a function of the electromagnetic field, the 
velocity of matter, the temperature, and other physical and geometrical 
variables. Quite remarkably, all the earlier results obtained 
for linear electrodynamics remain also valid in the general case because 
whatever local relation $H(F)$ may exist, the relation between
the {\it jumps} of the field derivatives, according to (\ref{kappa}),
is always linear. 

If we use Maxwell's equations (\ref{me}), then (\ref{had1}) and 
(\ref{had2}) yield
\begin{equation}\label{4Dwave}
  {\epsilon}^{\,ijkl}\, q_{j}\,h_{kl}=0 \,,\qquad
  {\epsilon}^{\,ijkl}\, q_{j}\,f_{kl}=0\,.
\end{equation}
Let us introduce the analog of the conventional constitutive matrix
\begin{equation}
\chi^{ijkl} := {\frac 1 2}\,\epsilon^{ijmn}\,\kappa_{mn}{}^{kl} 
= {\frac {\partial {\cal H}{}^{ij}}{\partial F_{kl}}},\label{chi}
\end{equation}
where we denote ${\cal H}{}^{ij} := {\frac 1 2}\,\epsilon^{ijmn}\,H_{mn}$.
Similarly to $\kappa_{ij}{}^{kl}$, we will often call the tensor 
$\chi^{ijkl}$ the jump tensor density. 

Now, making use of (\ref{kappa}) and (\ref{chi}), we rewrite the system 
(\ref{4Dwave}) as
\begin{equation}\label{4Dwave1}
  {\chi}^{\,ijkl}\, q_{j}\,f_{kl}=0 \,,\qquad
  {\epsilon}^{\,ijkl}\, q_{j}\,f_{kl}=0\,.
\end{equation}
Solving the last equation by $f_{ij} = q_ia_j - q_ja_i$, we finally
reduce (\ref{4Dwave1})$_1$ to
\begin{equation}\label{4Dwave3}
  {\chi}{}^{\,ijkl}\,q_{j}q_ka_l=0 \,.
\end{equation}
This algebraic system has a nontrivial solution for $a_i$ only when the 
wave covectors satisfy a certain condition. The latter gives rise 
to our {\it covariant Fresnel equation} \cite{OFR00,nonsym} 
\begin{equation} \label{Fresnel}  
{\cal G}^{ijkl}(\chi)\,q_i q_j q_k q_l = 0 \,,
\end{equation}
with the fourth order Fresnel tensor density $\cal G$ of weight $+1$
defined by
\begin{equation}\label{G4}  
  {\cal G}^{ijkl}(\chi):=\frac{1}{4!}\,\hat{\epsilon}_{mnpq}\,
  \hat{\epsilon}_{rstu}\, {\chi}^{mnr(i}\, {\chi}^{j|ps|k}\,
  {\chi}^{l)qtu }\,.
\end{equation}
It is totally symmetric, ${\cal G}^{ijkl}(\chi)= {\cal
  G}^{(ijkl)}(\chi)$, and thus has  35 independent components.

\section{Nonlinear electrodynamics}

Let us denote the two independent electromagnetic invariants as
\begin{equation}
I_1 := F_{ij}F^{ij},\qquad I_2 := F_{ij}\,\widetilde{F}^{ij},\label{I12}
\end{equation}
where $\widetilde{F}^{ij} = {\frac 1 2}\,\eta^{ijkl}\,F_{kl}$ and 
$\eta^{ijkl}:=|g|^{-1/2}\epsilon^{ijkl}$.
The Hodge operator for the exterior forms is denoted by the star ${}^\ast$,
as usual; but we will use a tilde to denote the dual 2-tensors.    
We will not restrict ourselves to the case of Minkowski spacetime with
$g_{ij} = {\rm diag}(1, -1, -1, -1)$, but instead $g_{ij}$ will be 
considered as an arbitrary curved Lorentzian spacetime metric.

The class of nonlinear electrodynamics models we study are described, 
in general, by the Lagrangian 4-form 
\begin{equation}
V = L\,\eta,\qquad {\rm with}\qquad L = L(I_1, I_2).\label{Lnonlin}
\end{equation}
Here, as usual, $\eta$ is the 4-form of the spacetime volume. The 
electromagnetic excitation 2-form, which enters the Maxwell equation
(\ref{me}), is derived as the derivative of the Lagrangian form, 
$H = - \partial V/\partial F$. Explicitly, we then have the nonlinear 
constitutive law
\begin{equation}
H = 4\left(-\,L_1\,{}^*\!F + L_2\,F\right).\label{MaxLL1}
\end{equation}
We denote the partial derivatives of the Lagrangian function w.r.t. 
its arguments as 
\begin{equation}
L_a := {\frac {\partial L}{\partial I_a}},\qquad
L_{ab} := {\frac {\partial^2 L}{\partial I_a\partial I_b}},\qquad
a,b = 1,2.
\end{equation}
In accordance with (\ref{kappa}) and (\ref{chi}), the direct 
differentiation of (\ref{MaxLL1}) yields the jump tensor
\begin{equation}
\chi^{ijkl} = \sqrt{|g|}\left[k_1\,g^{i[k}g^{l]j} + k_{2}\,F^{ij}\,F^{kl} 
+ k_{3}\,\widetilde{F}^{ij}\,F^{kl} + k_{4}\,F^{ij}\,\widetilde{F}^{kl} 
+ k_{5}\,\widetilde{F}^{ij}\,\widetilde{F}^{kl} + 
k_6\,\eta^{ijkl}\right].\label{Xeff}
\end{equation}
The coefficients $k_A, A=1,\dots, 6$, are functions of the 
electromagnetic fields:
\begin{equation}\label{kA}
k_1 = 4 L_1,\quad k_2 = 8L_{11},\quad k_3 =k_4 = 
8L_{12},\quad k_5 = 8L_{22},\quad k_6 = 2L_2.
\end{equation}
The identifications (\ref{kA}) are derived 
for the nonlinear Lagrangian (\ref{Lnonlin}) from the constitutive law 
(\ref{MaxLL1}). However, in most computations below we will consider 
the most general case with unspecified arbitrary coefficients $k_A$. 
This may be useful if we want to study the nonlinear electrodynamics of 
a more general type, for instance, with the dissipation effects and/or 
in moving media. 

In general, the untwisted tensor density $\chi^{ijkl}(x)$ of weight $+1$ 
has 36 independent components. We can decompose it into irreducible pieces 
\cite{nonsym} with respect to the 6-dimensional (``bivector'') 
linear group as follows:
\begin{equation}\label{decomp'}
\chi^{ijkl}={}^{(1)}\chi^{ijkl} + {}^{(2)}\chi^{ijkl} 
+ {}^{(3)}\chi^{ijkl}\,.
\end{equation}
The irreducible pieces of $\chi$ are defined by
\begin{eqnarray}
{}^{(2)}\chi^{ijkl}&:=&\frac{1}{2}\left( \chi^{ijkl} - \chi^{klij}\right) = -
\,{}^{(2)}\chi^{klij}\,,\quad {}^{(3)}\chi^{ijkl}:=\chi^{[ijkl]}\,,
\label{chiirr15} \\
{}^{(1)} \chi^{ijkl}&:=& \chi^{ijkl} -\,{}^{(2)}\chi^{ijkl}
-\,{}^{(3)}\chi^{ijkl}={}^{(1)} \chi^{klij}\,.\label{chiirr2}
\end{eqnarray}
The irreducible pieces  ${}^{(1)}\chi$, ${}^{(2)}\chi$, and ${}^{(3)}\chi$ 
have 20, 15, and 1 independent components, respectively. 
The possible presence of an 
axion piece ${}^{(3)}\chi$ was first studied by Ni \cite{Ni}, whereas a 
constitutive law with an isotropic skewon ${}^{(2)}\chi$ was discussed by 
Nieves and Pal \cite{NP}.

In the Lagrangian models (\ref{Lnonlin}), the effective constitutive tensor 
is automatically {\it symmetric}, i.e. ${}^{(2)}\chi=0$, which follows 
from (\ref{kA}), since 
$k_3=k_4$. However, in general the jump tensor (\ref{Xeff}) has all
the three irreducible pieces: 
\begin{eqnarray}
{}^{(1)}\chi^{ijkl} &=& \sqrt{|g|}\left[ k_1\,g^{i[k}g^{l]j} 
+ k_{2}\,F^{ij}\,F^{kl} 
+ {\frac {(k_3 + k_4)} 2}\,\left(\widetilde{F}^{ij}\,F^{kl} 
+ F^{ij}\,\widetilde{F}^{kl}\right)\right. \nonumber\\
&& \left. +\, k_{5}\,\widetilde{F}^{ij}\,\widetilde{F}^{kl} - \frac{1}{12}
\,[(k_3 + k_4)I_1 + (k_5-k_2)I_2]\,\eta^{ijkl}\right],\label{X1eff}\\
{}^{(2)}\chi^{ijkl} &=& {\frac {(k_3 - k_4)} 2}\sqrt{|g|}\,
\left(\widetilde{F}^{ij}\,F^{kl} - F^{ij}\,\widetilde{F}^{kl}\right),
\label{X2eff}\\
{}^{(3)}\chi^{ijkl} &=& {\frac 1 {12}}\sqrt{|g|}\left[(k_3 + k_4)I_1 
+ (k_5-k_2)I_2 + 12k_6\right]\eta^{ijkl}.\label{X3eff}
\end{eqnarray}

\section{Fresnel equation and birefringence}

Our study of the algebraic system in the framework of the Hadamard 
formalism yields the Fresnel equation in the generally covariant form
(\ref{Fresnel}) with the Fresnel tensor density (\ref{G4}). For the
explicit jump tensor density (\ref{Xeff}), it thus remains to substitute 
its components into (\ref{G4}). A straightforward calculation yields 
the result:
\begin{equation}
{\cal G}^{ijkl} = -\,{\frac {k_1} 8}\sqrt{|g|}\left({\cal X}\,g^{(ij}g^{kl)} 
+ 2{\cal Y}\,g^{(ij}t^{kl)} + {\cal Z}\,t^{(ij}t^{kl)}\right).\label{Gfin}
\end{equation}
Here we denote
\begin{equation}
t^{ij} := F^{ik}\,F^j{}_k,
\end{equation}
and 
\begin{eqnarray}
{\cal X} &=& k_1^2 + {\frac {k_1} 2}\,(k_3 + k_4)\,I_2 
- k_1k_5\,I_1 + {\frac 1 4}\,(k_3k_4 - k_2k_5)\,I_2^2,\label{Xfin}\\ 
{\cal Y} &=& k_1\,(k_2 + k_5) + (k_3k_4 - k_2k_5)\,I_1,\label{Yfin}\\ 
{\cal Z} &=& 4\,(k_2k_5 - k_3k_4)\,.\label{Zfin}
\end{eqnarray}

The most remarkable property of (\ref{Gfin}) is that it is obviously
factorizable into a product of 2 second order tensors. Correspondingly,
the quartic Fresnel surface of the wave normals reduces to the product
of two second order surfaces:
\begin{equation}\label{2cones}
{\cal G}^{ijkl}(\chi)\,q_i q_j q_k q_l = {\frac {-k_1} {8{\cal X}}}
\,(g^{ij}_{1}\,q_i q_j)\,(g^{kl}_{2}\,q_k q_l) = {\frac {-k_1} {8{\cal Z}}}
\,(\overline{g}{}^{ij}_{1}\,q_i q_j)\,(\overline{g}{}^{kl}_{2}\,q_k q_l).
\end{equation}
In other words, the wave normals lie not on the quartic surface but on
one of the two cones which are determined by the pair of {\it optical 
metric} tensors:
\begin{eqnarray}
g^{ij}_{1} &:=& {\cal X}\,g^{ij} + ({\cal Y} 
+ \sqrt{{\cal Y}^2 - {\cal XZ}})\,t^{ij},\label{omet1}\\
g^{ij}_{2} &:=& {\cal X}\,g^{ij} + ({\cal Y} 
- \sqrt{{\cal Y}^2 - {\cal XZ}})\,t^{ij}.\label{omet2}
\end{eqnarray}
The second equality in (\ref{2cones}) offers a different description of
the cones by means of the conformally equivalent metric tensors:
\begin{eqnarray}
\overline{g}{}^{ij}_{1} &:=& ({\cal Y} -\sqrt{{\cal Y}^2 - {\cal XZ}})
g^{ij} + {\cal Z}t^{ij} = {\frac 1 {\cal X}}({\cal Y} - \sqrt{{\cal Y}^2 
- {\cal XZ}})\,g^{ij}_{1},\label{omet1a}\\
\overline{g}{}^{ij}_{2} &:=& ({\cal Y} + \sqrt{{\cal Y}^2 - {\cal XZ}})
g^{ij} + {\cal Z}t^{ij} = {\frac 1 {\cal X}}({\cal Y} + \sqrt{{\cal Y}^2 
- {\cal XZ}})\,g^{ij}_{2}.\label{omet2a}
\end{eqnarray}
Thus, the general Fresnel analysis demonstrates that in {\em any} nonlinear 
electrodynamics model (\ref{Lnonlin}) the quartic wave surface {\it always} 
reduces to two light cones. This is the birefringence effect which
is thus a general feature of the nonlinear electrodynamics.

\section{Properties of optical metrics}

Let us discuss the results obtained in the previous section. 
The following general observations are in order. 

The Fresnel equation is trivially satisfied {\it for all} wave 
covectors when $k_1 =0$, see (\ref{Gfin}). Thus, -- in order to have waves
-- every electrodynamical Lagrangian $L$ {\it should necessarily} depend 
on the invariant $I_1 = F_{ij}F^{ij}$ (thus providing $k_1\neq 0$).

Accordingly, we will always assume that $k_1\neq 0$. 

In order to have a decent light propagation, the optical metrics should
be real and with Lorentzian signature. How can one be a priori sure 
that for every $L$ an optical metric necessarily has these properties? 

Using (\ref{Xfin})-(\ref{Zfin}) we find an explicit expression for the 
quantity under the square root in the above formulas :
\begin{equation}
{\cal Y}^2 - {\cal XZ} = N_1^2 + N_2\,N_3,\label{XYZ}
\end{equation}
where we have denoted
\begin{eqnarray}
N_1 &:=& k_1\,(k_2 - k_5) + (k_3k_4 - k_2k_5)\,I_1,\label{N1}\\
N_2 &:=& 2k_1k_3 + (k_3k_4 - k_2k_5)\,I_2,\label{N2}\\
N_3 &:=& 2k_1k_4 + (k_3k_4 - k_2k_5)\,I_2.\label{N3}
\end{eqnarray}
The expression (\ref{XYZ}) is {\it always non-negative} in every nonlinear 
theory (\ref{Lnonlin}) because $N_2=N_3$ when we take into account that 
$k_3 = k_4$, see (\ref{kA}). 

The signature of a four-dimensional metric is Lorentzian if and only if
its determinant is negative. Straightforward computation yields for
the optical metrics (\ref{omet1})-(\ref{omet2}):
\begin{equation}\label{det}
\left(\det{g^{ij}_a}\right) = \left(\det{g^{ij}}\right)\left[\alpha^2 + 
\frac{\alpha\beta_a}{2}\,I_1 + {\rm sign}(g)\frac{\beta_a^2}{16}\,I_2^2
\right]^2,\qquad a=1,2.
\end{equation}
Here $\alpha={\cal X}$ and $\beta_1 = {\cal Y} + \sqrt{{\cal Y}^2 - 
{\cal XZ}}$, $\beta_2 = {\cal Y} - \sqrt{{\cal Y}^2 - {\cal XZ}}$. As we 
see, both optical metrics have Lorentzian signature as soon as the 
spacetime metric $g^{ij}$ is Lorentzian. 

Summarizing, we have demonstrated that (\ref{omet1})-(\ref{omet2}) indeed 
describe the generic effect of a birefringent light propagation for {\it all} 
nonlinear Lagrangians. 

Recently, the emergence of the two ``effective geometries'' has been
described in \cite{nov1,nov2} without using the Fresnel approach. This
result is in a qualitative agreement with our analysis. In order to prove 
the quantitative correspondence, one needs to show that our optical metrics 
are conformally equivalent to the effective metrics of \cite{nov1,nov2}. 
Although the corresponding comparison is rather complicated and the direct 
proof is still not available, one can verify that $\sqrt{{\cal Y}^2 - 
{\cal XZ}}$ is equal to $64\sqrt{\Delta}$ of \cite{nov1}. Moreover, one
can recast the optical metrics (\ref{omet1})-(\ref{omet2a}) into the form
of the so-called Boillat metrics of \cite{GiHe}.

\section{Special Lagrangians}

It is worthwhile to study in a greater detail certain particular 
nonlinear models which are potentially of physical interest.

\subsection{Lagrangian $L=L(I_2)$} 

When the Lagrangian depends only on the second electromagnetic invariant,
we have $k_1 =0$, and there are no waves in such models.

\subsection{Lagrangian $L=L(I_1)$}

For the Lagrangian which, on the contrary, depends on the first invariant
only, we find $k_3 =k_4 =k_5= 0$. Accordingly, from (\ref{Xfin})-(\ref{Zfin}) 
we find ${\cal X} = k_1^2,\ {\cal Y}= k_1k_2,\ {\cal Z}=0$ and thus 
(\ref{omet1})-(\ref{omet2}) yield
\begin{equation}
g^{ij}_{1} = k_1\left(k_1\,g^{ij} + 2k_2\,t^{ij}\right),\qquad 
g^{ij}_{2} = k_1^2\,g^{ij}.\label{ordinary}
\end{equation}
Correspondingly, we still have birefringence with some photons moving
along the standard null rays of the spacetime metric $g^{ij}$, whereas
other photons choosing the rays null with respect to the optical metric 
$L_1\,g^{ij} + 4L_{11}\,t^{ij}$, cf. \cite{nov1,nov2}.

\subsection{Lagrangian $L=U(I_1) + \alpha\,I_2$}

This is a simple generalization of the above case. Here $\alpha$ does not
depend on the electromagnetic field, although it is not a constant, in
general. When it depends on the spacetime coordinates, $\alpha=\alpha(x)$, 
one can identify it with the axion field. 

Here we again have $k_3 =k_4 =k_5= 0$ and we recover the same light
cone structure (\ref{ordinary}). To put it differently, axion does not
disturb the light cones which are solely determined by the spacetime
metric and by the dependence of the Lagrangian on the invariant $I_1$.

\subsection{Lagrangian $L=a\,I_1 + V(I_2)$}

Then $k_2 =k_3 =k_4= 0$ which yields ${\cal X} = k_1^2 - k_1k_5\,I_1,
\ {\cal Y}= k_1k_5,\ {\cal Z}=0$. Consequently, 
\begin{equation}
g^{ij}_{1} = k_1\left[(k_1 - k_5\,I_1)\,g^{ij} + 2k_5\,t^{ij}\right],
\qquad g^{ij}_{2} = k_1(k_1 - k_5\,I_1)\,g^{ij},
\end{equation}
i.e., there is again birefringence with one cone determined by the 
standard spacetime metric.

\subsection{Born-Infeld theory}

The Lagrangian of the Born-Infeld (BI) theory \cite{BI} reads:
\begin{equation}
L = b^2\left(\sqrt{1 + {\frac 1 {2b^2}}\,I_1 
- {\frac 1 {16b^4}}\,I_2^2} - 1\right).\label{BIlagr}
\end{equation}
Here $b$ is the coupling constant. By differentiation, we find:
\begin{eqnarray}
k_1 &=& 4L_1 = {\frac 1 {\left({1 + {\frac 1 {2b^2}}
\,I_1 - {\frac 1 {16b^4}}\,I_2^2}\right)^{1/2}}},\\
k_2 &=& 8L_{11} = {\frac {-1} {2b^2\left({1 + {\frac 1 {2b^2}}
\,I_1 - {\frac 1 {16b^4}}\,I_2^2}\right)^{3/2}}},\\
k_3 &=& k_4 = 8L_{12} = {\frac {I_2} {8b^4\left({1 + {\frac 1 {2b^2}}
\,I_1 - {\frac 1 {16b^4}}\,I_2^2}\right)^{3/2}}},\\ 
k_5 &=& 8L_{22} = {\frac {-\left(1 + {\frac 1 {2b^2}}\,I_1\right)} 
{2b^2\left({1 + {\frac 1 {2b^2}}\,I_1 - {\frac 1 {16b^4}}
\,I_2^2}\right)^{3/2}}}.
\end{eqnarray}
Correspondingly,
\begin{eqnarray}
{\cal X} &=& {\frac {\left(1 + {\frac 1 {2b^2}}\,I_1\right)^2} 
{\left({1 + {\frac 1 {2b^2}}\,I_1 - {\frac 1 {16b^4}}
\,I_2^2}\right)^{2}}},\\
{\cal Y} &=& {\frac {-\left(1 + {\frac 1 {2b^2}}\,I_1\right)} 
{b^2\left({1 + {\frac 1 {2b^2}}\,I_1 - {\frac 1 {16b^4}}
\,I_2^2}\right)^{2}}},\\
{\cal Z} &=& {\frac 1 {b^4\left({1 + {\frac 1 {2b^2}}\,I_1 
- {\frac 1 {16b^4}}\,I_2^2}\right)^{2}}}.
\end{eqnarray}
As a result, we find ${\cal Y}^2 - {\cal XZ} = 0$, and thus the two
optical metrics (\ref{omet1})-(\ref{omet2}) coincide,
\begin{equation}
g^{ij}_{1} = g^{ij}_{2} = {\frac {\left(1 + {\frac 1 {2b^2}}\,I_1\right)} 
{\left({1 + {\frac 1 {2b^2}}\,I_1 - {\frac 1 {16b^4}}\,I_2^2}\right)^2}}
\left[\left(1 + {\frac 1 {2b^2}}\,I_1\right)g^{ij} 
- {\frac 1 {b^2}}t^{ij}\right].
\end{equation}
Birefringence disappears, and the photons propagate along a single
light cone determined by the ``quasi-metric'' of Plebanski \cite{pleb}
\begin{equation}
\left(1 + {\frac 1 {2b^2}}\,I_1\right)g^{ij} - {\frac 1 {b^2}}t^{ij}. 
\end{equation}

\section{No birefringence condition}

In this section, we will restrict our attention to the Lagrangian 
theories for which the constitutive tensor is (\ref{Xeff}), and the 
coefficients are derived as (\ref{kA}). It is important that the Fresnel 
analysis reveals that $k_1\neq 0$, otherwise there is no decent wave 
propagation at all. 

As it is clear from (\ref{XYZ}), the necessary and sufficient 
condition of the absence of birefringence is provided by the pair
of equations:
\begin{eqnarray}
N_1 &=& k_1\,(k_2 - k_5) + (k_3k_4 - k_2k_5)\,I_1 = 0,\label{nobr1}\\
N_2 &=& N_3 = 2k_1k_3 + (k_3k_4 - k_2k_5)\,I_2 = 0.\label{nobr2}
\end{eqnarray}
Here the property $k_3=k_4$ of the Lagrangian models is used. 

Taking into account that all $k$'s are the partial derivatives of the 
Lagrangian $L$ w.r.t. $I_1$ and/or $I_2$ as displayed in (\ref{kA}), we can 
view the above system as a pair of partial differential equations, the 
solution $L=L(I_1, I_2)$ of which describes a {\it model without 
birefringence} (i.e., with a single light cone). At least two such
particular solutions are already known: one is rather simple, namely, the 
standard Maxwell theory with $L = I_1/4$. Another is more nontrivial -- 
this is the Born-Infeld theory with the Lagrangian (\ref{BIlagr}). 
One may ask the question: Are these the only solutions of the system 
(\ref{nobr1})-(\ref{nobr2})? The immediate inspection of the system
(\ref{nobr1})-(\ref{nobr2}) shows that the answer is negative. For
example, the Lagrangian function $L(I_1, I_2) = aI_1/I_2$, with 
constant $a$, satisfies the equations (\ref{nobr1})-(\ref{nobr2}). Such 
a nonlinear (and nonpolynomial) model thus also has no birefringence. 
It is an open problem to find the complete set of solutions of 
(\ref{nobr1})-(\ref{nobr2}), leading then to a single light cone.

\section{Closure condition}

Let us denote the ``traceless'' part of the jump tensor as 
\begin{equation}
\not\!\chi^{ijkl} := {}^{(1)}\!\chi^{ijkl} + {}^{(2)}\!\chi^{ijkl}.
\end{equation}
As we know \cite{nonsym}, only the traceless part determines 
the Fresnel surface, whereas the axion part ${}^{(3)}\!\chi^{ijkl}$ drops 
out completely from the wave propagation analysis. 

For the general jump tensor (\ref{Xeff}), we find:
\begin{eqnarray}
\frac{1}{4}\,\hat{\epsilon}_{pqij}\not\!\chi^{ijkl}\,\hat{\epsilon}_{klrs}
\not\!\chi^{rsmn} &=& \left(-\,k_1^2 + {\frac {a_0^2}{k_1^2}}\right)
\delta^{[m}_p\delta^{n]}_q  + a_0\,\eta_{pq}{}^{mn} \nonumber\\
&& +\,a_1\,F_{pq}\widetilde{F}^{mn} 
+ a_2\,\widetilde{F}_{pq}F^{mn} + \,a_3\,F_{pq}F^{mn}  
+ a_4\,\widetilde{F}_{pq}\widetilde{F}^{mn}.\label{clo}
\end{eqnarray}
Here the coefficients read: 
\begin{eqnarray}
a_0 &=& {\frac {k_1} 6}\left[(k_3 + k_4)\,I_1 + (k_5 - k_2)\,I_2\right] = 
{\frac 16}\left[-\,N_1\,I_2 + {\frac {(N_2 + N_3)}2}\,I_1\right],\label{a0}\\
a_1 &=& -\,(k_3 + k_4)\,k_1 - k_3k_4\,I_2 + {\frac {k_5} 3}\left[
2(k_3 + k_4)\,I_1 + (2k_5 + k_2)\,I_2\right] \nonumber\\
&=& N_1\left(-\,{\frac {2k_5}{3k_1}}\,I_2\right) + {\frac {(N_2 + N_3)} 2}
\left(-\,1 + {\frac {2k_5}{3k_1}}\,I_1\right),\label{a1}\\
a_2 &=& -\,(k_3 + k_4)\,k_1 - k_3k_4\,I_2 + {\frac {k_2} 3}\left[
2(k_3 + k_4)\,I_1 + (k_5 + 2k_2)\,I_2\right] \nonumber\\
&=& N_1\left({\frac {2k_2}{3k_1}}\,I_2\right) + {\frac {(N_2 + N_3)} 2}
\left(-\,1 - {\frac {2k_2}{3k_1}}\,I_1\right),\label{a2}\\
a_3 &=& (k_5 - k_2)\,k_1 + k_2k_5\,I_1 + {\frac {k_3} 3}\left[
2(k_5 - k_2)\,I_2 + (2k_3 - k_4)\,I_1\right]\nonumber\\
&=& N_1\left(-\,1 -{\frac {2k_3}{3k_1}}\,I_2\right) + {\frac {(N_2 + N_3)} 2}
\left({\frac {2k_3}{3k_1}}\,I_1\right),\label{a3}\\
a_4 &=& -\,(k_5 - k_2)\,k_1 - k_2k_5\,I_1 - {\frac {k_4} 3}\left[
2(k_5 - k_2)\,I_2 + (2k_4 - k_3)\,I_1\right]\nonumber\\
&=& N_1\left(1 + {\frac {2k_4}{3k_1}}\,I_2\right) + {\frac {(N_2 + N_3)} 2}
\left(-\,{\frac {2k_4}{3k_1}}\,I_1\right).\label{a4}
\end{eqnarray}
The second lines in (\ref{a1})-(\ref{a4}) give the $a$'s in terms of the 
combinations (\ref{nobr1})-(\ref{nobr2}). Certainly, we use the assumption 
that $k_1\neq 0$. 

Like the constitutive tensor of linear electrodynamics, the jump 
tensor $\not\!\kappa_{ij}{}^{kl}${}$= {\frac 1 2}\,\hat{\epsilon}_{ijmn}$
{\,}$\not\!\chi^{mnkl}$ determines a linear map in the 6-dimensional space 
of 2-forms. When the action of this map, repeated twice, brings us (up to a 
factor) back 
to the identity map, we speak of the {\it closure} property of 
$\not\!\kappa_{ij}{}^{kl}$. The importance of the closure property is 
related to the fact that ultimately $\not\!\kappa_{ij}{}^{kl}$ turns out
to the {\it duality operator} which determines a unique conformal Lorentzian
metric on the spacetime. 

In nonlinear electrodynamics, the jump tensor (\ref{Xeff}) 
has the closure property when 
\begin{equation}
a_0 =0,\qquad a_1 = 0,\qquad a_2 =0,\qquad a_3 =0,\qquad a_4 =0,\label{condC}
\end{equation}
as is evident from (\ref{clo}).

\section{Equivalence of closure and no birefringence conditions}

In linear electrodynamics, there is much evidence (although
the final rigorous proof is still missing) that the quartic Fresnel
surface of wave covectors reduces to a unique light cone if and only
if the constitutive tensor has the closure property.

For the nonlinear Lagrangian theories it is possible to make some 
progress in solving the equivalence problem. In a certain sense, the
situation here is simpler because we have discovered that the quartic
Fresnel surface is always reduced to the product of the light
cones (birefringence). The next step is thus to study under which 
conditions the birefringence disappears and, correspondingly, a 
unique light cone arises.
\medskip

\noindent {\it Theorem}:
In the general nonlinear electrodynamical model described by the Lagrangian 
(\ref{Lnonlin}), the Fresnel equation implies a single light 
cone (no birefringence) if and only if the traceless part of the jump tensor 
satisfies the closure property. 
\medskip

\noindent {\it Proof}: As a preliminary remark, we note that for the
Lagrangian models (\ref{Lnonlin}) the jump tensor (\ref{Xeff}) is 
symmetric because $k_3=k_4$. As a result, $N_2 = N_3$. 

The necessary condition is evident. The birefringence is absent when 
$N_1=N_2(=N_3)=0$, see (\ref{nobr2}). Then we immediately read from 
(\ref{a0})-(\ref{a4}) that $a_0 =a_1 =a_2 =a_3 = a_4 =0$, and thus 
the closure is recovered from (\ref{clo}). 

The sufficient condition is also proved straightforwardly. The closure
condition (\ref{condC}) has the unique solution
\begin{equation}
N_1 =0,\qquad {\frac {N_2 + N_3} 2} =0,\label{condC2}
\end{equation}
when we analyze (\ref{a0})-(\ref{a4}). Since for the Lagrangian models we 
have $N_2=N_3$, then (\ref{condC2}) yields $N_1=N_2=N_3 =0$. Thus, there 
is no birefringence. 

To put it differently, we have proven that the closure of the traceless
jump tensor is the necessary and sufficient condition for the reduction 
of the fourth order Fresnel wave surface to a single light cone. This is 
true for {\it all nonlinear} Lagrangian electrodynamical theories 
(\ref{Lnonlin}). Returning to our studies of the general {\it linear} 
electrodynamics, we expect that a similar result holds true there.

\subsection{On asymmetric jump tensors}

Symmetry of the jump tensor is very important in the equivalence proof
above. In order to clarify this point, let us consider an arbitrary
jump tensor (\ref{Xeff}) without assuming the explicit form of the
coefficients (\ref{kA}). When $k_3\neq k_4$, the jump tensor has the
nontrivial skewon part (\ref{X2eff}). Also, $N_2\neq N_3$ [in fact, as 
we can see from (\ref{N2}) and (\ref{N3}), $N_2 - N_3 = 2k_1(k_3 -k_4)$].

We can easily verify that the closure of an asymmetric operator is not 
equivalent to the no-birefringence property. Indeed, take, for instance, 
$k_1\neq 0$, $k_2=k_3=0$, $k_4\neq 0$ and $k_5= 0$. Then the jump tensor 
is asymmetric and $N_1=N_2 =0$ but $N_3\neq 0$. We then 
obtain a unique light cone because (\ref{XYZ}) vanishes identically. However,
the closure condition (\ref{condC}) is not satisfied, since
(\ref{a0})-(\ref{a4}) are nontrivial for $N_1=N_2=0$ and $N_3\neq 0$. 
This also means that the requirement of a unique light cone does not 
necessarily implies that $\chi$ must be symmetric.

The opposite is also true: Suppose an asymmetric jump tensor (\ref{Xeff})
has the closure property, i.e. (\ref{condC}) is fulfilled. Then we find
(\ref{condC2}) again. However, (\ref{condC2}b) yields $N_2 = - N_3$,
and consequently (\ref{XYZ}) together with (\ref{omet1}) and (\ref{omet2})
describe the case of a birefringent and dissipative wave propagation.

These examples show that the closure of an {\it asymmetric} jump (or 
constitutive) tensor does not guarantee the absence of birefringence, 
and, vice versa, no-birefringence is not accompanied by the closure 
property for an asymmetric operator. 


\section{Moving isotropic nonlinear media}

Recently, there has been some interest in the light propagating in 
moving media with nontrivial dielectric and magnetic properties. The
first covariant analysis of the Fresnel equation for this case was
done by Kremer \cite{kremer}. An isotropic medium is characterized
by the constitutive law
\begin{equation}
{\cal H}^{ij} = \sqrt{|g|}\,\left[{\frac 1 \mu}\,F^{ij} + 2\left(
{\frac 1 \mu} - \varepsilon\right) u_k\,F^{k[i}\,u^{j]}\right],\label{movCL}
\end{equation}
where $u^i$ is the 4-velocity of the moving matter (normalized as
usual by $u_iu^i = 1$), and $\varepsilon$ and $\mu$ are the permeability 
and permittivity functions of the isotropic medium. 
The case when they do not depend on the electromagnetic field strength 
(being constant in space and time, for example) was investigated in 
\cite{kremer}. 

More recently, the nonlinear case when $\varepsilon = \varepsilon(F)$ 
and $\mu =\mu(F)$ are functions of the electromagnetic field has been 
studied by De Lorenci et al \cite{nov2}. However, the attention was 
restricted to certain special cases, and the general result is still missing.

We can perform a fairly complete analysis of the wave propagation
in a nonlinear moving media on the basis of our covariant Fresnel
equation (\ref{Fresnel}). By differentiation, we easily find the jump
tensor (\ref{chi}):
\begin{eqnarray}
\chi^{ijkl} &=& \sqrt{|g|}\,\Bigg[{\frac 2 \mu}\,g^{i[k}g^{l]j} + 
4\left({\frac 1 \mu} - \varepsilon\right) u^{[k}\,g^{l][i}\,u^{j]} 
\nonumber\\ && +\,m^{kl}\,F^{ij} + 2\left(m^{kl} -
e^{kl}\right) u_m\,F^{m[i}\,u^{j]}\Bigg].\label{chimov}
\end{eqnarray}
The second line is absent in the linear theory, and the tensors
\begin{equation}
m^{ij} := {\frac {\partial (1/\mu)}{\partial F_{ij}}},\qquad
e^{ij} := {\frac {\partial \varepsilon}{\partial F_{ij}}},
\end{equation}
are responsible for the nonlinear electrodynamical effects.

Inspection immediately reveals that the jump tensor density (\ref{chimov})
contains {\it both} an axion and a skewon part. There are claims in the
literature that axion and skewon, in general, do not have physical sense. 
However, here we encounter a simple and a physically sound example with the 
both parts being nontrivial. The irreducible pieces 
(\ref{chiirr15})-(\ref{chiirr2}) read: 
\begin{eqnarray}
{}^{(1)}\chi^{ijkl}&=&\sqrt{|g|}\left[{\frac 2 \mu}g^{i[k}g^{l]j}
+4\left({\frac 1 \mu}-\varepsilon\right)u^{[k}g^{l][i}u^{j]}
+\frac{1}{2}\left(m^{ij}F^{kl}+m^{kl}F^{ij}\right) 
+\left(m^{kl}-e^{kl}\right)P^{[i}u^{j]}\right. \nonumber \\
&&\left. \qquad +\left(m^{ij}-e^{ij}\right)P^{[k}u^{l]} 
+ {\frac 1 {12}}\,\eta^{ijkl}\left[F^{mn}\widetilde{m}_{mn} 
+ 2(\widetilde{m}_{mn} - \widetilde{e}_{mn})P^mu^n\right]\right],
\end{eqnarray} 
\begin{equation}
{}^{(2)}\chi^{ijkl}=\sqrt{|g|}\left[
\frac{1}{2}\left(F^{ij}m^{kl}-F^{kl}m^{ij}\right) 
+\left(m^{kl}-e^{kl}\right)P^{[i}u^{j]}
-\left(m^{ij}-e^{ij}\right)P^{[k}u^{l]}\right] ,
\end{equation}
and for the axion piece, we have
\begin{equation}
\hat\epsilon_{ijkl}\chi^{ijkl}=\sqrt{|g|}\hat\epsilon_{ijkl}\left[
F^{ij}m^{kl}+2\left( m^{kl}-e^{ij}\right) P^iu^j\right].
\end{equation}
Thus, nonlinear isotropic matter does have axion and skewon induced
by nonlinearity.

\subsection{Nonmagnetic matter}

Let us consider the case when the magnetic constant is independent
of the electromagnetic field, that is $m^{ij} =0$. [In the simplest
case, we can restrict the attention to the purely dielectric medium with
$\mu =1$. However, we will formally keep $\mu\neq 1$, for the sake of
generality]. 

It is convenient to make the evident split of (\ref{chimov}) into the 
sum $\chi^{ijkl} = \phi^{ijkl} + \psi^{ijkl}$, with
\begin{eqnarray}
\phi^{ijkl} &=& \sqrt{|g|}\,\left[{\frac 2 \mu}\,g^{i[k}g^{l]j} + 
4\left({\frac 1 \mu} - \varepsilon\right) u^{[k}
\,g^{l][i}\,u^{j]}\right],\label{phi1}\\
\psi^{ijkl} &=& -\,2\sqrt{|g|}\,e^{kl}\,u_m
\,F^{m[i}\,u^{j]}.\label{psi1}
\end{eqnarray}
It is straightforward to find the Fresnel tensor for the first piece:
\begin{equation}
{\cal G}^{ijkl}(\phi) = {\frac \varepsilon {\mu^2}}
\,{\rm sign}(g)\,\sqrt{|g|}\,\,{\stackrel{\rm o}{g}}{}^{(ij}
\,{\stackrel{\rm o}{g}}{}^{kl)}.\label{Gphi}
\end{equation}
Here we have denoted the so called {\it Gordon optical metric} \cite{Gordon} as
\begin{equation}
{\stackrel{\rm o}{g}}{}^{ij} := g^{ij} + (\varepsilon\mu - 1)\,u^iu^j.
\end{equation}
Its inverse reads
\begin{equation}
{\stackrel{\rm o}{g}}{}_{ij} := g_{ij} + \left({\frac 1 {\varepsilon\mu}} 
- 1\right)\,u_iu_j,
\end{equation}
and the determinant can be easily computed: $\det {\stackrel{\rm o}{g}}
= (\det g)/(\varepsilon\mu)$. 

Using this optical metric we can simplify the form of (\ref{phi1}), 
bringing it to
\begin{equation}
\phi^{ijkl} = {\frac 2 \mu}\,\sqrt{|g|}\,\,{\stackrel{\rm o}{g}}{}^{i[k}
{\stackrel{\rm o}{g}}{}^{l]j} = 2\,\sqrt{\frac \varepsilon \mu}
\,\sqrt{|{\stackrel{\rm o}{g}}|}\,\,{\stackrel{\rm o}{g}}{}^{i[k}
{\stackrel{\rm o}{g}}{}^{l]j}.\label{phi2}
\end{equation}

Now, for $\chi = \phi + \psi$, using a compact notation by omitting 
the indices, we have 
\begin{equation} \label{over1}
{\cal G}(\chi) = {\cal G}(\phi) + {\cal G}(\psi) 
+ {\frac 1{4!}}\left(O_1 + O_2 + O_3 + T_1 + T_2 + T_3\right).
\end{equation}
Here the mixed terms $O_a$ contain one $\psi$-factor and the $T_a$'s
two $\psi$-factors. {\it Postponing the symmetrization} over $i,j,k,l$
to the very last moment, these terms read explicitly as follows:
\begin{eqnarray}
O_1(\phi,\psi,\phi) &=& \hat{\epsilon}_{mnpq}\,\hat{\epsilon}_{rstu}\, 
\phi^{mnri}\,\psi^{jpsk}\,\phi^{lqtu}\,,\label{o1}\\
O_2(\psi,\phi,\phi) &=& \hat{\epsilon}_{mnpq}\,\hat{\epsilon}_{rstu}\, 
\psi^{mnri}\,\phi^{jpsk}\,\phi^{lqtu}\,,\label{o2}\\
O_3(\phi,\phi,\psi) &=& \hat{\epsilon}_{mnpq}\,\hat{\epsilon}_{rstu}\,
\phi^{mnri}\,\phi^{jpsk}\,\psi^{lqtu}\,,\label{o3}\\ 
T_1(\psi,\phi,\psi) &=& \hat{\epsilon}_{mnpq}\,\hat{\epsilon}_{rstu}\, 
\psi^{mnri}\,\phi^{jpsk}\,\psi^{lqtu}\,,\label{t1}\\
T_2(\psi,\psi,\phi) &=& \hat{\epsilon}_{mnpq}\,\hat{\epsilon}_{rstu}\, 
\psi^{mnri}\,\psi^{jpsk}\,\phi^{lqtu}\,,\label{t2}\\
T_3(\phi,\psi,\psi) &=& \hat{\epsilon}_{mnpq}\,\hat{\epsilon}_{rstu}\,
\phi^{mnri}\,\psi^{jpsk}\,\psi^{lqtu}\,.\label{t3}
\end{eqnarray}

An important observation is that all $T$'s are vanishing for (\ref{psi1}).
Indeed, let us denote $P^j := u_i\,F^{ij}$, then 
\begin{equation}\label{psi2}
\psi^{ijkl} = -\,2\,\sqrt{|g|}\,P^{[i}\,u^{j]}\,e^{kl}\,.
\end{equation}
Then we straightforwardly find, for example,
\begin{eqnarray}
T_1(\psi,\chi,\psi) &=& 4|g|\,\hat{\epsilon}_{mnpq}\,\hat{\epsilon}_{rstu}\,
P^{[m}\,u^{n]}e^{ri}\,P^{[l}\,u^{q]}e^{tu}\,
\chi^{jpsk}\nonumber\\ &=& 2|g|\,\hat{\epsilon}_{mnpq}\,\hat{\epsilon}_{rstu}
\,\chi^{jpsk}\,P^{m}\,u^{n}(P^{l}\,u^{q} - P^{q}\,u^{l})
\,e^{ri}\,e^{sk} =0.\label{T1}
\end{eqnarray}
This is zero because either the symmetric $u^nu^q$ or symmetric 
$P^mP^q$ is contracted with the antisymmetric $\hat{\epsilon}_{mnpq}$.
Note that we on purpose write $\chi^{jpsk}$ as the second argument,
because its form is {\it arbitrary}, not necessarily equal to 
(\ref{phi1}). Analogously, we find:
\begin{eqnarray}
T_2(\psi,\psi,\chi) &=& 4|g|\,\hat{\epsilon}_{mnpq}\,\hat{\epsilon}_{rstu}
\,P^{[m}\,u^{n]}e^{ri}\,P^{[j}\,u^{p]}e^{sk}\,
\chi^{lqtu}\nonumber\\ &=& 2|g|\,\hat{\epsilon}_{mnpq}\,\hat{\epsilon}_{rstu}
\,\chi^{lqtu}\,P^{m}\,u^{n}(P^{j}\,u^{p} - P^{p}\,u^{j})
\,e^{ri}\,e^{sk} =0.\label{T2}
\end{eqnarray}
It is a little bit more nontrivial to prove that $T_3$ also vanishes.
We have, explicitly:
\begin{eqnarray}
T_3(\chi,\psi,\psi) &=& 4|g|\,\hat{\epsilon}_{mnpq}\,\hat{\epsilon}_{rstu}
\,P^{[j}\,u^{p]}e^{sk}\,P^{[l}\,u^{q]}e^{tu}\,
\chi^{mnri}\nonumber\\ &=& |g|\,\hat{\epsilon}_{mnpq}\,\hat{\epsilon}_{rstu}
\,\chi^{mnri}\,(P^{j}\,u^{p}P^{l}\,u^{q} - P^{p}\,u^{j}P^{l}\,u^{q}
\nonumber\\ && \qquad -\,P^{j}\,u^{p}P^{q}\,u^{l} 
+ P^{p}\,u^{j}P^{q}\,u^{l})\,e^{sk}\,e^{tu}
=0.\label{T3}
\end{eqnarray}
The first and the last terms in the parentheses contain the 
symmetric combinations $u^pu^q$ and $P^pP^q$ which are vanishing
when contracted with the antisymmetric $\hat{\epsilon}_{mnpq}$. The
two remaining terms in the parentheses are reduced, by means
of a relabeling of indices, to $-\,P^{p}\,u^{q}(P^{l}\,u^{j} 
- P^{j}\,u^{l})$. Recalling that at the end we impose the
symmetrization over the free indices $(i,j,k,l)$, we thus 
prove that $T_3(\chi,\psi,\psi) =0$. 

Since in all the three formulas (\ref{T1})-(\ref{T3}), the argument
$\chi^{jpsk}$ is completely arbitrary, we can put $\chi^{jpsk} =
\psi^{jpsk}$, in particular. Then (\ref{T1})-(\ref{T3}) yields
that ${\cal G}(\psi) =0$. As the next choice, we put $\chi^{jpsk} =
\phi^{jpsk}$. Then (\ref{over1}) combined with (\ref{T1})-(\ref{T3}), 
yields 
\begin{equation}\label{over2}
{\cal G}(\chi) = {\cal G}(\phi) + {\frac 1{4!}}\left(O_1 + O_2 + O_3\right). 
\end{equation}
It thus remains to compute the $O$-terms. The corresponding calculation
is straightforward and simple if we use the representation (\ref{phi2}). 
Then we find:
\begin{equation}
O_1(\phi,\psi,\phi) = O_2(\psi,\phi,\phi) = O_3(\phi,\phi,\psi)
= -\,8\,{\rm sign}(g)\,{\frac \varepsilon \mu}\,
\,\,{\stackrel{\rm o}{g}}{}^{(ij}\,\psi^{k|mn|l)}
\,{\stackrel{\rm o}{g}}{}_{mn}.\label{O123}
\end{equation}
Note that this result is valid {\it for all} possible tensors
$\psi$, not only (\ref{psi1}) which means that we can further
use (\ref{O123}) for the future calculations involving more general
nonlinear pieces (in particular, for the case with a nontrivial
$m^{ij}$). Using then (\ref{O123}) in (\ref{over2}), we get
\begin{equation}
{\cal G}^{ijkl}(\phi + \psi) =  {\frac \varepsilon {\mu^2}}
\,{\rm sign}(g)\,\sqrt{|g|}\,\,{\stackrel{\rm o}{g}}{}^{(ij}
\,{\stackrel{\rm 1}{g}}{}^{kl)}.\label{Gchi}
\end{equation}
Here we denoted
\begin{equation}
{\stackrel{\rm 1}{g}}{}^{ij} := {\stackrel{\rm o}{g}}{}^{ij} 
- {\frac \mu {\sqrt{|g|}}}\,\psi^{(i|mn|j)}
\,{\stackrel{\rm o}{g}}{}_{mn}.\label{opt2}
\end{equation}
Hence, a purely dielectric nonlinear moving medium will in general
exhibit the birefringence effect: the light will propagate in such
a medium along the cone of the original optical metric 
${\stackrel{\rm o}{g}}{}^{ij}$ (one may call it ordinary ray), and along 
the second cone determined by the metric ${\stackrel{\rm 1}{g}}{}^{ij}$
(``extraordinary'' ray). 

The explicit form of the extraordinary optical metric is obtained when
we substitute (\ref{psi2}) into (\ref{opt2}):
\begin{eqnarray}
{\stackrel{\rm 1}{g}}{}^{ij} &=& {\stackrel{\rm o}{g}}{}^{ij} 
- \,\mu P_k \,e^{k(i}\,u^{j)} + {\frac 1 \varepsilon}
\,u_ke^{k(i}\,P^{j)},\label{opt3}\\
&=& {\stackrel{\rm o}{g}}{}^{ij} + \,\mu\,\,e^{k(i}\,u^{j)}F_{kl}u^l 
- {\frac 1 \varepsilon}\,u_ke^{k(i}\,F^{j)l}u_l.\label{opt35}
\end{eqnarray}
Special cases of this general result were considered in \cite{nov2}.

As an example, let us consider a medium in its rest frame. 
We use adapted coordinates such that 
$u^i = \delta^i_0 = \left(1,\ \vec{0}\right)$. We additionally assume that 
the dielectric permittivity is given by 
\begin{equation}
\varepsilon = \overline{\varepsilon} + a\,\vec{E}^2.\label{optKerr1}
\end{equation}
Here $\overline{\varepsilon}$ and $a$ are constant parameters. The
components of the electric vector are defined as usual, $\vec{E} =
E_a = -\,F_{0a}$. Then we find 
\begin{equation}
e^{0a} = -\,e^{a0} = -\,2a\,E^a,\qquad {\rm and}
\qquad P^j = u_i\,F^{ij} = \left(0,\ E^b\right).\label{optKerr2}
\end{equation}
The spatial indices are lowered and raised with the help of the
3-metric $\delta_{ab}$ (we neglect the gravitational). 
As a result, from (\ref{optKerr2}) we obtain
\begin{equation}
P_k\,e^{ki} = \left(-\,2a\,\vec{E}^2,\ \vec{0}\right),\qquad
u_k\,e^{ki} = \left(0,\ -\,2a\,E^b\right).\label{optKerr3}
\end{equation}
Finally, using all this in (\ref{opt3}) we find the components of the
extraordinary optical metric:
\begin{equation}\label{optKerr4}
g^{00} = n^2\left(1 + {\frac {2a} \varepsilon}\,\vec{E}^2\right),
\qquad g^{ab} = -\,\delta^{ab} - {\frac {2a} \varepsilon}\,E^aE^b.
\end{equation}
Here $n:=\sqrt{\varepsilon\mu}$ is the refraction index of the medium. 
In this way we obtain a natural description of the optical Kerr effect
(see \cite{LL60,LL84}, for example) when birefringence is induced by
the applied electric field.

\section{Discussion and conclusion}

In this paper, we have performed a systematic analysis of the light 
propagation in the nonlinear electrodynamics on the basis of the Fresnel
approach. We have considered two classes of models: (a) general 
nonlinear Lagrangian theories, and (b) moving nonlinear matter. In the
former case, the Lagrangian (\ref{Lnonlin}) is an arbitrary function
of the two electromagnetic invariants, whereas in the latter case, the
permeability and permittivity functions of the medium (\ref{movCL}) depend 
arbitrarily on the electromagnetic field.

The study of the first class of models reveals the {\it generic} nature of 
the birefringence effect: The quartic Fresnel surface reduces to the
two light cones for {\it all} nonlinear Lagrangians. We show that the
resulting optical metrics are always real and have the correct Lorentzian
signature. In this way, we confirm and extend the recent results of
\cite{nov1,nov2}. Furthermore, we are able to demonstrate the validity 
of the so called closure--no birefringence conjecture in the context of 
nonlinear electrodynamics: Birefringence is absent (and thus the 
quartic Fresnel 
surface reduces to a unique light cone) if and only if the effective
constitutive (or jump) tensor satisfies closure property. 

The nonlinear moving matter with the constitutive law (\ref{movCL}) gives
a sound example of a model in which the effective constitutive tensor
naturally has nontrivial axion and skewon contributions. Accordingly,
one should then expect that the Fresnel surface remains quartic, in 
general \cite{nonsym}. However, for nonmagnetic material
media, we show that birefringence is again the generic effect. The
Fresnel surface factorizes into two light cones, one of which 
corresponds to the Gordon optical metric (independent of nonlinearities),
whereas the other (\ref{opt3}) manifests the nonlinear properties of the
model. The optical Kerr effect represents a particular example of our 
general derivations.

\bigskip
{\bf Acknowledgments}.
The work of YNO was partially supported by FAPESP, and GFR would like to 
thank the German Academic Exchange Service (DAAD) for financial support.

\end{document}